\documentclass{PoS}

\usepackage{subfigure}
\usepackage{amsmath}

\makeatletter

\providecommand{\tabularnewline}{\\}
\newcommand{\lyxdot}{.}

\usepackage{slashed}

\newcommand{\DSoneH}{{\cal H}^{(\Delta S=1)}}

\def\mpi2{m_\pi^2}
\def\mK2{m_K^2}

\def\mres{m_{\rm res}}

\newcommand{\bea}{\begin{eqnarray}}
\newcommand{\eea}{\end{eqnarray}}
\newcommand{\be}{\begin{equation}}
\newcommand{\ee}{\end{equation}}

\def\lvec#1{\setbox0=\hbox{$#1$}
    \setbox1=\hbox{$\scriptstyle\leftarrow$}
    #1\kern-\wd0\smash{
    \raise\ht0\hbox{$\raise1pt\hbox{$\scriptstyle\leftarrow$}$}}
    \kern-\wd1\kern\wd0}
\def\rvec#1{\setbox0=\hbox{$#1$}
    \setbox1=\hbox{$\scriptstyle\rightarrow$}
    #1\kern-\wd0\smash{
    \raise\ht0\hbox{$\raise1pt\hbox{$\scriptstyle\rightarrow$}$}}
    \kern-\wd1\kern\wd0}

\makeatother

\title{Kaon Weak Matrix Elements in 2+1 flavor DWF QCD}

\ShortTitle{Kaon Weak Matrix Elements in 2+1 flavor DWF QCD}

\author{\speaker{Shu Li} \\
       Department of Physics, Columbia University, New York, NY
       10027, USA \\
       E-mail: \email{lishu@phys.columbia.edu}}

\author{\speaker{Robert D. Mawhinney} \\
        Department of Physics, Columbia University, New York, NY
	10027, USA \\
        E-mail: \email{rdm@physics.columbia.edu}}

\author{RBC and UKQCD Collaborations}

\abstract{
$K \to \pi$ and $K \to 0$ weak matrix elements of $\Delta S = 1$
operators have been measured in 2+1 flavor domain wall fermion (DWF)
QCD on (3 fm)$^3$ lattices with $a^{-1} = 1.73(3)$ GeV.
As is well known, using these
matrix elements and chiral perturbation theory allows a determination
of the $K \to \pi \pi$ matrix elements that enter in the quantitative
value for the $\Delta I = 1/2$ rule and $\epsilon^\prime/\epsilon$.
Two light dynamical sea quark masses have been used, along 
with six valence quark masses, with the lightest valence quark mass
$\approx 1/10$ the physical strange quark mass.  We report our
results for lattice matrix elements in the $SU(3)_L \times SU(3)_R$
(27,1), (8,1), and (8,8) representations, paying particular attention
to the statistical errors achieved after measurements on 75
configurations.  We also report on our calculation of the
non-perturbative renormalization coefficients for these $\Delta
S=1$ weak operators, using the Rome-Southampton method.}

\FullConference{The XXV International Symposium on Lattice Field Theory\\
                July 30 - August 4 2007\\
                Regensburg, Germany}

\begin{document}

\section{Introduction}

CP violation is now a well established property of Nature, having
been observed in both kaon and B meson systems.  Its standard model
explanation, that the violation arises from the difference between
the mass and weak interaction eigenstates of the quarks, is
quantitatively encapsulated in the unitary $3 \times 3$
Cabbibo-Kobayashi-Maskawa (CKM) matrix which relates these eigenstates.
Determining the elements of this matrix is the subject of extensive
theoretical and experimental effort.  However, even assuming precise
values for the elements of the CKM matrix, a prediction for the
rate of direct CP-violating decays in kaons, determined experimentally
by measuring the quantity Re($\epsilon^\prime/\epsilon$), is lacking.
Such a prediction relies on values for the hadronic $K \to \pi \pi$
matrix elements of low-energy, four-quark operators.  Lattice QCD
is the only first principles method to allow determination of these
matrix elements and their measurement has been the subject of many
studies.  In this note, we discuss our progress in determining these
matrix elements, using full 2+1 flavor QCD with domain wall quarks.

Euclidean space lattice methods for the direct evaluation of the
desired $K \to \pi \pi$ matrix elements have been developed, and
some preliminary studies have been performed
\cite{Lellouch:2000pv,Kim:2004sk}.  However, to reach physical
values for the kaon and pion masses in full QCD simulations is
beyond the reach of current computers.  Here we follow the long-standing
approach of using chiral perturbation theory to relate $K \to \pi
\pi$ matrix elements to matrix elements of the same operator measured
in $K \to \pi $ and $K \to 0$ matrix elements \cite{Bernard:1985wf}.
Quenched calculations have been done with this approach
\cite{Blum:2001xb,Noaki:2001un}, which showed that with DWF, the
problems of lower dimensional operator mixing at finite lattice
spacing and renormalization of the operators were under good control.
Statistically well resolved values for the lowest order chiral
perturbation theory constants were achieved and, by naively extending
lowest order chiral perturbation theory to the kaon mass, physical
results were quoted.

However, for these particular kaon weak matrix elements calculations,
it was pointed out that quenching makes the relation between
the full QCD chiral perturbation theory constants and the quenched
ones ambiguous \cite{Golterman:2001qj,Golterman:2002us,Golterman:2006ed}, 
providing a source of uncertainty in
interpreting the quenched calculation that is more serious than
the general caution that must be exercised with quenching.  In the
current 2+1 flavor calculation, these ambiguities are removed.

To date, our calculation involves measuring the same operators as
in our previous quenched calculation \cite{Blum:2001xb}, although
on larger volume lattices, with 2+1 dynamical flavors and with much
lighter valence quark masses \cite{Antonio:2006zza,Antonio:2007pb}.
With these valence masses, we see
that next-to-leading order (NLO) SU(3) chiral perturbation theory
for pseudoscalar decay constants and masses \cite{Lin:2007lat},
and the pseudoscalar bag parameter $B_{PS}$ (the analogue of $B_K$
with two light quarks in the ``kaon") \cite{Antonio:2007lat}, represents
the lattice data at the $\approx 5$\% level or better.  Thus we can
try similar NLO chiral perturbation theory fits to the data presented
here, which covers a similar range of pseudoscalar masses, in the
hopes of extracting the desired low energy constants.  These fits
require the 2+1 flavor partially quenched chiral perturbation theory
formula for our matrix elements, a calculation which is underway
\cite{Aubin:2007xpt}.

We also discuss the status of our calculation of the non-perturbative
renormalization Z factors needed to convert our lattice results
to $\overline{\rm MS}$ conventions.

\section{Simulation Details}

\begin{figure}
\begin{centering}
\includegraphics[scale=0.5]{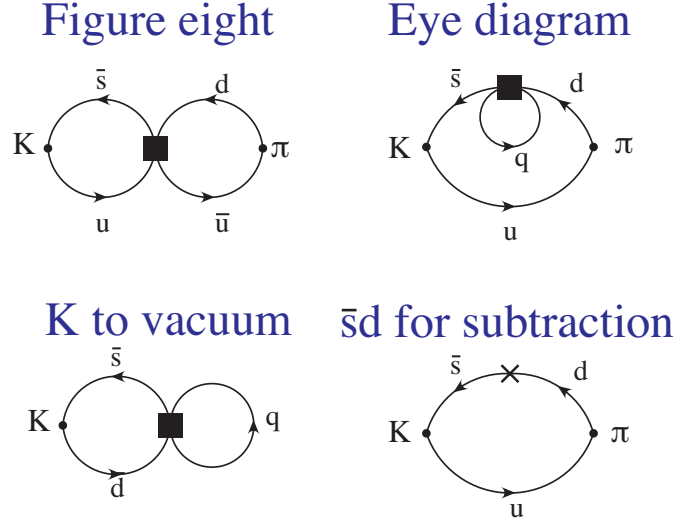}
\caption{The contractions used in the evaluation of the $\Delta S = 1$
operators that have been calculated in this report.  The sources for the
kaons and pions are Coulomb gauge fixed wall sources and the closed
fermion loop contractions are calculated with stochastic sources.
\label{fig:contractions}}
\end{centering}
\end{figure}

We have measured the 10 $\Delta S = 1$ operators needed for a
determination of $\epsilon^\prime/\epsilon$ and the $\Delta I =
1/2$ rule, using the notation of \cite{Blum:2001xb}.  When evaluated
between $K^+$ and $\pi^+$ states, the operators produce the
contractions shown in Figure \ref{fig:contractions}. The contractions
between kaons and pions are known as figure eight diagrams and eye
diagrams.  We also need vacuum diagrams and contractions involving
$\bar{s}d$, to remove the mixing with lower dimensional operators.
We use Coulomb gauge fixed wall sources at $t = 5$ and 59 on our
$24^3 \times 64 \times 16$ lattices as the sources for the kaons
and pions.  The Dirac equation is solved twice for each wall source,
once with periodic and once with anti-periodic boundary conditions,
and the two solutions added to effectively double the length of the
lattice in the time direction.  Our operator contractions are always
done with $5 \leq t_{op} \leq 59$, where $t_{op}$ is the time where
the operator is evaluated, so the double lattice merely serves to
suppress any contributions to our operators from propagation around
the world in the time direction.

For the closed fermion loops in the eye and $K \to 0$ contractions,
we use a stochastic source, spread over the spatial volume and
covering all time slices from 12 to 51 (40 time slices).  A single
stochastic source is a fixed spin and color component, with a
space-time value that is Gaussianly distributed.  For each spin and
color combination, solutions are found for four different stochastic
space-time distributions.

For each configuration, 6 valence quark masses are used: $m_f =
0.001$, 0.005, 0.01, 0.02, 0.03, 0.04.  The residual quark mass
from the finite $L_s$ extent of our lattices is 0.00315(2)
\cite{Lin:2007lat}.  These bare quark masses correspond to
pseudoscalar masses ranging from 250 MeV to 750 MeV.  From fits to
pseudoscalar masses and decay constants, we find that for the three
lightest bare quark masses (0.001, 0.005 and 0.01), NLO chiral
perturbation theory agrees well with our measured data.

For each configuration, we do 8 full (all spins and colors for the
source) five-dimensional solutions to the Dirac equations for each
valence mass.  Each wall source requires 2 solutions (one for each
boundary condition) and we do four stochastic estimators for the
closed fermion loops.  Thus we spend half of the time on the wall
source inversions and half on the stochastic inversions.

The results reported here are for 76 configurations, separated by
40 molecular dynamics time units, for our ensemble with light dynamical
quark mass $m_l = 0.005$ and 74 configurations, with the same
separation, for $m_l = 0.01$.  To do these 8 full inversions for
the six valence quark masses on a single configuration takes about
24 hours on a 4,096 node QCDOC partition, with the lightest valence
quark mass taking about half of the total time.

\section{Chiral Perturbation Theory and Ward Identity Test}

In the 3 flavor effective theory that comes from integrating out the
charm, bottom and top quarks (as well as the weak vector bosons),
the effective Hamiltonian for $\Delta S = 1$ transitions is
\begin{equation}
     \DSoneH =
     \frac{G_F}{\sqrt{2}} V_{ud}V_{us}^* \left\{
     \sum_{i=1}^{10} \left[ z_i(\mu) + \tau y_i(\mu) \right] Q_i
     \right\}
\end{equation}
Here $z_i(\mu)$ and $y_i(\mu)$ are Wilson coefficients and the $Q_i$
are four-quark operators.  Of the ten four-quark operators $Q_i$
above, only seven are independent, and they include three
representations of $SU(3)_L \times SU(3)_R$: (27,1), (8,1) and
(8,8).  In lowest order chiral perturbation theory, the desired $K
\to \pi \pi$ matrix elements are all proportional to an appropriate
low energy constant, denoted as $\alpha_1^{(8,1)}$, $\alpha^{(27,1)}$,
or $\alpha^{(8,8)}$.  These come from $K \to \pi$ and $K \to 0$
matrix elements as given in Table \ref{tab:chpt}.  Note that for
the (8,1) operators, to determine the physically relevant quantity
$\alpha_1^{(8,1)}$ requires remove the power divergent quantity
$\alpha_2^{(8,1)}$, which is measured in $K \to 0$ matrix elements

\begin{table}
\begin{center}
  \begin{tabular}{c|c|l|c|c} \hline
   Irrep & Number & Isospin
    & $K^+ \to \pi^+$ &  $K^0 \to \pi^+ \pi^-$  \\ \hline
  (27,1) &   1	& 1/2, 3/2
     & $ -\frac{4 m_M^2}{f^2} { \alpha^{(27,1)}}$
  	 & $ -\frac{4 i}{f^3} m_{K^0}^2
  	      { \alpha^{(27,1)}}$  \\ \hline
  (8,1)  &   4	& 1/2
    & $  \frac{4 m_M^2}{f^2} ({ \alpha^{(8,1)}_1}
  	     - { \alpha^{(8,1)}_2})$
  	 & $  \frac{4 i}{f^3} m_{K^0}^2
  	      { \alpha^{(8,1)}_1}$  \\ \hline
  (8,8)  &   2	& 1/2, 3/2
      & $ -\frac{12}{f^2} { \alpha^{(8,8)}}$
  	 & $ -\frac{12 i}{f^3} { \alpha^{(8,8)}}$
  	 \\ \hline
  \end{tabular}
  \end{center}
\caption{The relation between matrix elements and low energy constants,
to lowest order in chiral perturbation theory. The second column
gives the number of different operators transforming as the
given irreducible representation.
\label{tab:chpt}}
\end{table}

To accurately determine the leading order chiral perturbation theory
constants, we want to try to fit our matrix element data to the
full partially quenched 2+1 flavor formulae.  Calculations of these
formulae are underway \cite{Aubin:2007xpt}, and when they are complete, we
will fit our data to them in our efforts to extract the best values
for the $\alpha$'s.

Since NLO chiral perturbation theory plays such an important role
here, we turn briefly to a simple matrix element $ \langle \pi^+ |
\bar{s} d | K^+ \rangle$ that is known in chiral perturbation theory
and that also enters our calculation.  At leading order in chiral
perturbation theory, one has
\begin{equation}
 \frac{2 m_f}{m_\pi^2} \langle \pi^+ | \bar{s} d | K^+ \rangle 
     - 1 = 0
\end{equation}
We calculate this from the ratio of a three point function to two,
2-point functions as
\begin{equation}
  R_2 \equiv
      \frac{\langle P_{\pi^+}^{\rm wall}(x_0) \, [\bar{s} d](y) \,
         P_{K^-}^{\rm wall}(z_0) \rangle }
	{ \langle P_{\pi^+}^{\rm wall}(x_0) \, P_{\pi^-}^{\rm wall}(z_0)
	   \rangle}
\end{equation}
Here $P^{\rm wall}_{\pi^+}$ is a wall Coulomb gauge fixed pseudoscalar
wall source with the quantum numbers of the $\pi^+$.  $R_2$ differs
from $ \langle \pi^+ | \bar{s} d | K^+ \rangle$ by known factors.

\begin{figure}
\begin{centering}
\includegraphics[scale=0.5]{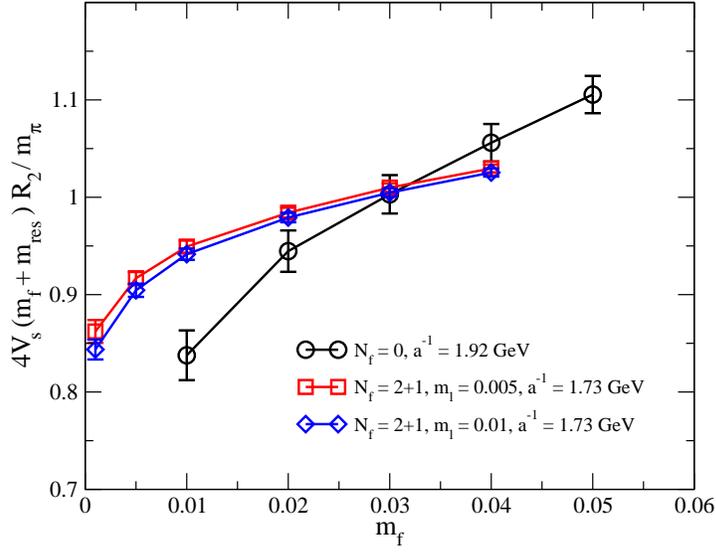}
\caption{A plot of the appropriately scaled $\langle \pi^+ |
\bar{s} d | K^+ \rangle $ versus valence quark mass.  When the valence
and sea quark masses are equal, the curves should go to 1.}
\label{fig:ward_sbard.eps}
\end{centering}
\end{figure}

Figure \ref{fig:ward_sbard.eps} shows a plot of the appropriately
scaled $R_2$ versus valence input quark mass for the current 2+1
flavor full QCD ensembles and the same results for our earlier
quenched ensemble.  In the limit $m_f \to -\mres$ and the light sea
and valence quarks are equal, the curves should go to 1.  We can
see that there is noticeable curvature in the graphs for quark
masses below 0.01 and there is a small dynamical quark mass effect
when comparing the $m_l = 0.005$ and 0.01 ensembles.  This is
certainly the region where partially quenched chiral logarithms
should be noticeable, and are seen in our other observables, but
fits to the precise formulae (when available) will be required to
sharpen this general statement.  For $B_{PS}$ \cite{Antonio:2007lat} we
do see that the partially quenched logarithms have a curvature that
is opposite the curvature when the valence and sea quark masses are the
same.  Also note that the curvature is much more noticeable in the
dynamical ensemble data than the quenched results and that the valence
masses in the current work are much lighter than the earlier
quenched work.

\section{$\Delta I = 3/2$ Matrix Elements}

\begin{figure}
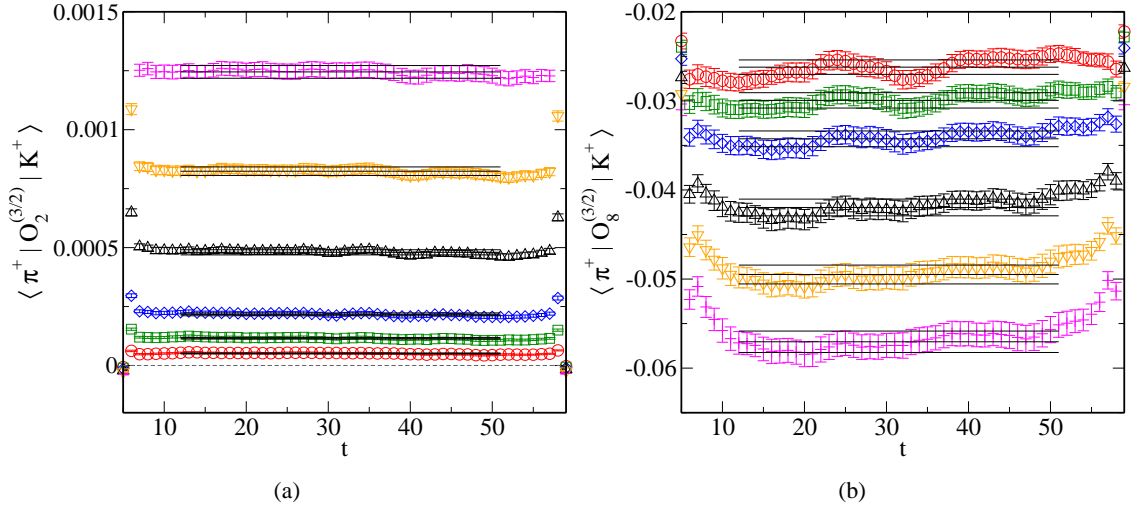

\begin{centering}
\subfigure[]{\includegraphics[clip,scale=0.6]
{fig/plateau32_wall_pseudo/O2_plat_deg.eps}}
\subfigure[]{\includegraphics[clip,scale=0.6]
{fig/plateau32_wall_pseudo/O8_plat_deg.eps}}
\caption{Plateaus for the $\Delta I = 3/2$ part of the $K \to \pi$
matrix elements of $Q_2$ and $Q_8$.  For $Q_2$ (left panel) the
valence masses run from the largest (top points) to the smallest
(bottom), while for $Q_8$ (right panel) they run from the smallest
(top points) to the largest (bottom).  The pseudoscalars have
degenerate quark masses.  \label{fig:plateau32}}
\end{centering}
\end{figure}

We now turn to the $\Delta I = 3/2$ matrix elements and present some
of our lattice data.  The 3/2 part of $Q_2$ and $Q_8$ are presented in
Figure \ref{fig:plateau32}.  We see at once that we have very long
plateaus which extend, with essentially uniform errors, across the
lattice.  We choose to fit the plateau from $t=12$ to 51, {\em i.e.}
starting from a distance of 7 away from the two wall sources.  Clearly
with such long plateaus the results are not sensitive to the precise
fit range chosen.  The average value, with error bars, for each mass is
indicated by the horizontal lines on the graphs.

\begin{figure}
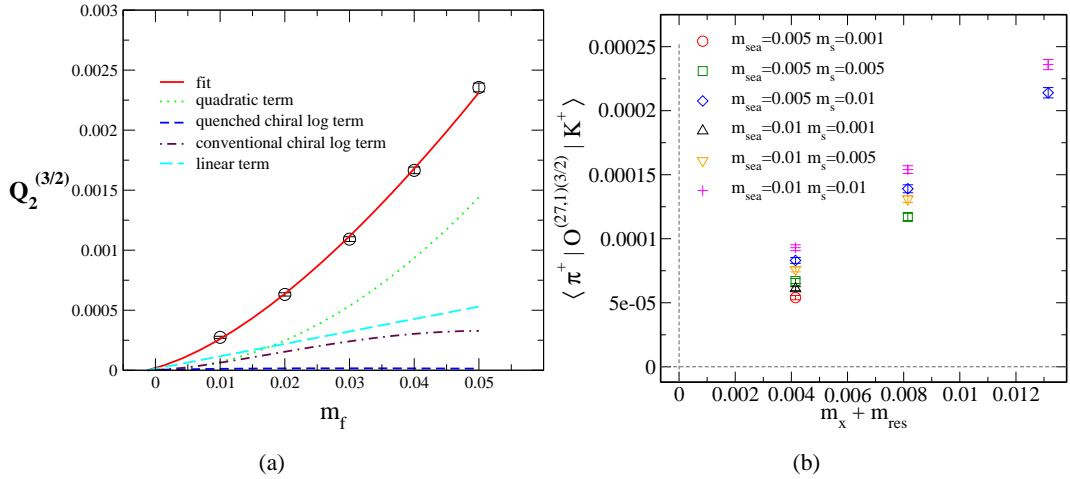

\begin{centering}
\subfigure[]{\includegraphics[clip,scale=0.37]
{fig/theta271_32_corr_mpi.eps}}
\subfigure[]{\includegraphics[clip,scale=0.55]
{fig/O27_all/O27_32_pseudo5/alpha27_NLO.eps}}
\caption{A comparison of the lattice normalized matrix elements for
our earlier quenched calculation (left panel) and the current
2+1 flavor calculation (right panel).
\label{fig:matrix_elements_32}}
\end{centering}
\end{figure}

In Figure \ref{fig:matrix_elements_32}, we present values for
$Q_2^{(3,2)}$ as a function of valence quark mass for our previous
quenched calculation (left panel) and the $m_l = 0.005$ lattices
from our current ensemble.  Note the much smaller range of quark
masses plotted in the 2+1 flavor QCD case.  In our earlier quenched
work, we saw no strong indication of quenched chiral logarithms,
but found a large contribution from the analytic NLO terms.  We do
not yet have the chiral perturbation theory formulae for the 2+1
flavor case, nor the fits to the data.  However, one can see that some
downward curvature is needed to have the extrapolated matrix element
vanish at the origin, as required by chiral symmetry.  Also, for these
much smaller valence quark masses, one does not readily see a
noticeable effect of higher order analytic terms.

\section{Weak Matrix Elements in the (8,1) Representation}

\subsection{(8,1) Operators}

Among the 10 weak operators involved in the $K\to\pi\pi$ process
below the charm threshold, $Q_{i}\,\left\{ i=3,4,5,6\right\} $ are
pure (8,1) $\Delta I=1/2$ operators, while $Q_{i}\,\left\{
i=1,2,9,10\right\} $ have both (8,1), $\Delta I=1/2$ and (27,1),
$\Delta I=1/2$ parts.  The (8,1) operators are particularly important
for the investigation of CP violation, since, from the size of
the Wilson coefficients at 2 GeV, the matrix element of $Q_{6}$
is dominant in $\mathrm{Im}\left(A_{0}\right)$ \cite{Blum:2001xb}.

Since the $\Delta I=1/2$ operators can mix with the quadratically
divergent operator $\Theta^{ \left(3,\bar{3}\right)} \equiv \bar{s}
\left(1-\gamma_{5} \right)d$, their $K\to\pi$ matrix elements alone
cannot determine the relevant $K\to\pi\pi$ elements. It is
necessary to evaluate their $K\to 0$ matrix elements to remove
the unwanted contributions.

\subsection{Plateaus of $K\to\pi$ for (8,1) Operators}

\begin{figure}
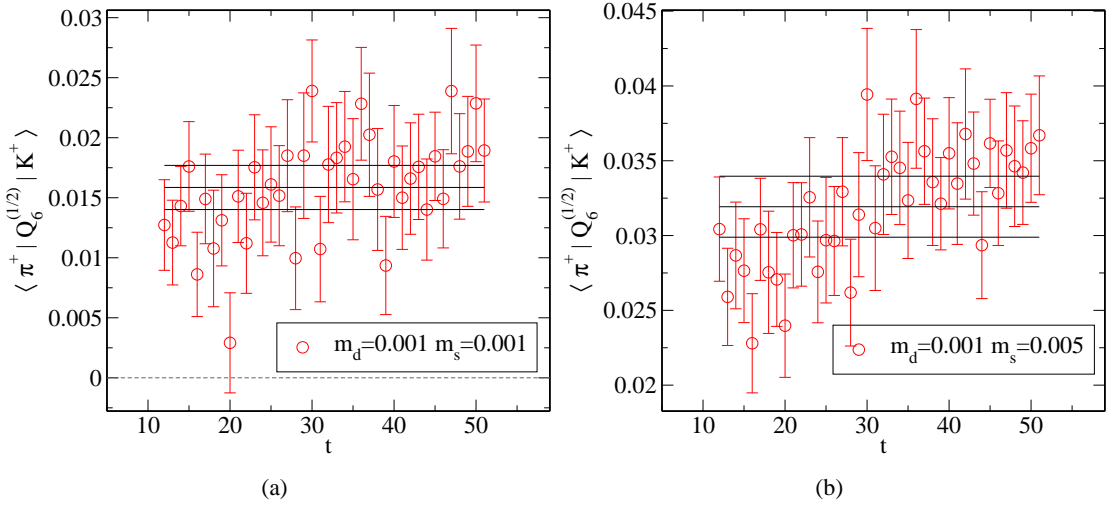

\begin{centering}
\subfigure[]{\includegraphics[clip,scale=0.6]
{raw/O6_re_plat_md0\lyxdot 001_ms0\lyxdot 001}}
\subfigure[]{\includegraphics[clip,scale=0.6]
{raw/O6_re_plat_md0\lyxdot 001_ms0\lyxdot 005}}
\par\end{centering}
\caption{Plateaus of $\left\langle
\pi^{+}\left|Q_{6}\right|K^{+}\right\rangle $ with
$m_{\mathrm{sea}}=0.005$. Figure (a) has degenerate valence masses
($m_{d}=m_{s}=0.001$) and Figure b) has non-degenerate valence quark
($m_{d}=0.001$, $m_{s}=0.005$). In both cases, the matrix elements
have long plateaus. \label{fig:Q6 K->pi plat}}
\end{figure}

As mentioned earlier, we evaluate the value of $\left\langle \pi^{+}
\left| Q_i \right| K^{+}\right \rangle $, by placing Coulomb gauge
fixed wall sources at $t_{K}=5$ and $t_{\pi}=59$.   The eye contractions
that appear in (8,1) operators are determined with random sources
with support from $t_{rand}=12$ to 51.  We can then evaluate the
operator contribution from any timeslice given by $t_{op}$ with $12
\leq t_{op} \leq 51$.  By taking the ratios of the three- and
two-point pseudoscalar-pseudoscalar Green's functions we calculate
the required matrix elements \cite{Blum:2001xb}.  As an example,
Figure \ref{fig:Q6 K->pi plat} shows the plateaus for $\left\langle
\pi^{+}\left|Q_{6}\right|K^{+}\right\rangle $. With either degenerate
or non-degenerate valence masses, we have very long plateaus (40
time-slices).  Comparing these graphs with the plateaus in Figure
\ref{fig:plateau32}, we see that the error bars here, after averaging
over time slices (black horizontal lines) are smaller than the error
bars for the individual points.  Such an effect is not very noticeable
in the $\Delta I =3/2$ amplitudes of Figure \ref{fig:plateau32}.

\subsection{Resolving Quadratic Divergence with $K\to 0$ Matrix
Elements}

\begin{figure}
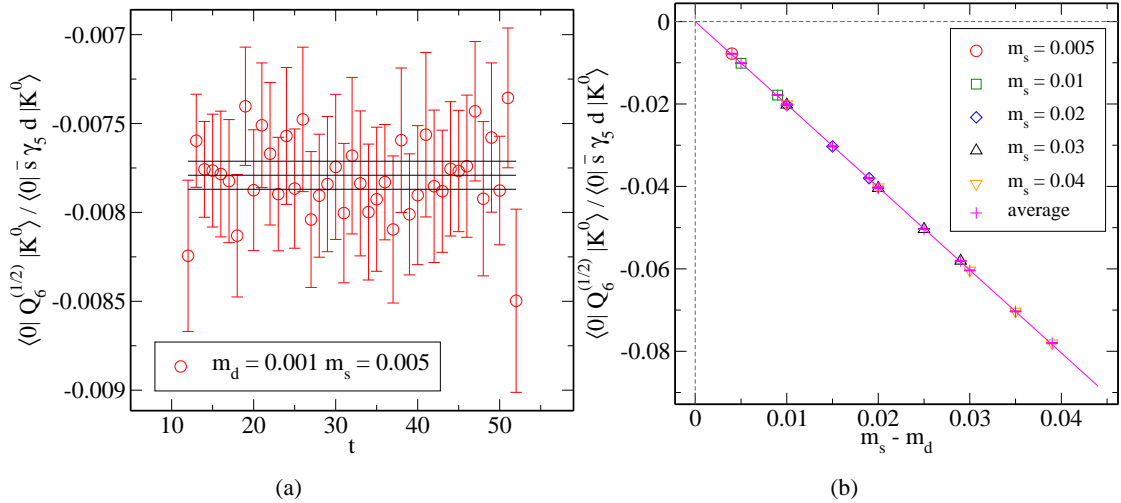

\begin{centering}
\subfigure[]{\includegraphics[clip,scale=0.6]{raw/K6_re_msea0\lyxdot 005_md0\lyxdot 001_ms0\lyxdot 005}}\subfigure[]{\includegraphics[clip,scale=0.6]{raw/K6_re_msea0\lyxdot 005_kvac_fit}}
\par\end{centering}
\caption{Calculating mixing coefficient between four-quark operators and $\bar{s}d$
using the ratio $\left\langle 0\left|Q_{i,\mathrm{lat}}^{\left(1/2\right)}\right|K^{0}\right\rangle /\left\langle 0\left|\left(\bar{s}\gamma_{5}d\right)_{\mathrm{lat}}\right|K^{0}\right\rangle $.
Figure (a) shows the plateau of the matrix element with $m_{\mathrm{sea}}=0.005$,
$m_{d}=0.001$, and $m_{s}=0.005$. Figure (b) shows the results of fitting to $(m_s-m_d)$,
which reveals a nice linear structure.\label{fig:Q6 K->0 plat fit}}
\end{figure}

As explained earlier, we need to calculate the $K\to 0$ matrix
elements for $\Delta I=1/2$ operators to remove the mixing with
lower dimensional operators. For the (8,1) operators, we determine
a mixing coefficient $\eta_{1,i}$ as in \cite{Blum:2001xb} by
calculating
\begin{equation}
\frac{\left\langle 0\left|Q_{i,\mathrm{lat}}^{
\left(1/2\right)}\right|K^{0}\right\rangle }
{\left\langle 0\left|
\left(\bar{s}\gamma_{5}d\right)_{\mathrm{lat}}
\right|K^{0}\right\rangle }
=\eta_{0,i}+\eta_{1,i}\left(m_{s}-m_{d}\right)
\label{eq:K->0 fit}
\end{equation}
where the valence quark masses have $m_{s}\neq m_{d}$. Again using
$Q_{6}$ as an example, Figure \ref{fig:Q6 K->0 plat fit} shows the
plateau of the $K\to 0$ matrix element in the left panel and the
result of fitting to Eq.\ \ref{eq:K->0 fit} above.  One sees that
the data shows a very accurate linear dependence on $m_s - m_d$, which
is expected since the quadratically divergent part of $Q_6$
dominates the numerator and it is proportional to $m_s - m_d$.  Since
$Q_{6}$ has the largest divergent contribution, its matrix elements
show the best linearity.  For other operators, with a smaller divergent
part, the linearity is less good, but also the required cancellation in
the subtraction of the divergent part is less stringent.

\subsection{Subtraction of the Quadratic Divergence}

After calculating the mixing coefficient $\eta_{1,i}$, we use it
to subtract the quadratically divergent contribution of $\left\langle \pi^{+}\left|\bar{s}d\right|K^{+}\right\rangle $
from the raw matrix elements \cite{Blum:2001xb}. \begin{align}
\left\langle \pi^{+}\left|Q_{i}^{\left(8,1\right)}\right|K^{+}\right\rangle _{\mathrm{sub}} & \equiv\left\langle \pi^{+}\left|Q_{i}^{\left(8,1\right)}\right|K^{+}\right\rangle +\eta_{1,i}\left(m_{s}+m_{d}\right)\left\langle \pi^{+}\left|\bar{s}d\right|K^{+}\right\rangle \label{eq:K->pi sub}\\
 & =\frac{4\alpha_{1}^{\left(8,1\right)}}{f^{2}}m_{K}m_{\pi}+\left(chiral\; logs\right)+(analytic\; terms)\label{eq:K->pi sub ChPT}\end{align}
where the second line is the prediction of the chiral perturbation
theory. (The cancellation of the power divergences is true without
reference to chiral perturbation theory.) While we have begun fitting
to next-to-leading order 2+1 flavor partially quenched chiral
perturbation theory, the formulae are still being checked, so
we have also experimented with simple leading order fits.

\begin{figure}
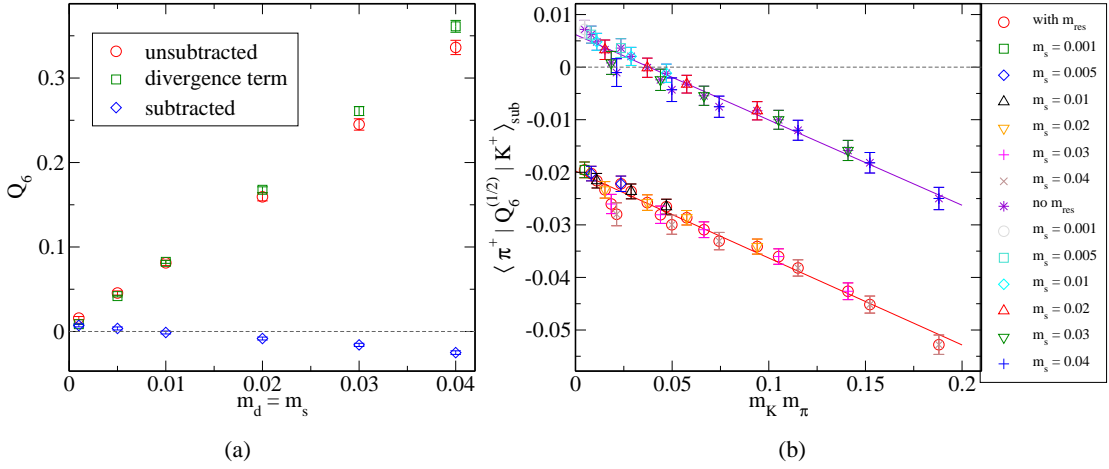

\begin{centering}
\subfigure[]{\includegraphics[clip,scale=0.55]{raw/O6_vacsub}}
\subfigure[]{\includegraphics[clip,scale=0.55]{raw/O6_sub_Ktopi12_fit_nondeg}}
\par\end{centering}

\caption{Subtraction of quadratic divergence from $K\to\pi$ matrix elements.
Figure (a) shows the subtraction process for degenerate valence masses
$m_{d}=m_{s}$. It is clear that for the divergent operator, the physical
quantity is very sensitive to any noise in the matrix elements. Figure (b)
fits the subtracted matrix elements $\left\langle \pi^{+}\left|Q_{6}\right|K^{+}\right\rangle _{\mathrm{sub}}$
with all available pairs of non-degenerate valence masses to the leading
order term $m_{K}m_{\pi}$. The upper line does not use $m_{\mathrm{res}}$
while on the lower line each subtraction has used $m_{\mathrm{res}}$.
Both figures present the data of $m_{\mathrm{sea}}=0.005$ ensemble.\label{fig:Q6 K->pi sub fit}}
\end{figure}

Figure \ref{fig:Q6 K->pi sub fit}(a) shows the subtraction process
for $Q_{6}$. Since $Q_{6}$ is the most divergent, the subtraction
term (green squares) is very close to the unsubtracted matrix
elements, and thus a precise determination of the subtraction term
is vital.  (This was also the case in the quenched calculations.)
In Figure \ref{fig:Q6 K->pi sub fit}(b), we show the subtracted
matrix elements $\left\langle \pi^{+}\left|Q_{6} \right|K^{+} \right
\rangle _{\mathrm{sub}}$, including all non-degenerate combinations
of valence masses $\left(m_{d},\, m_{s}\right)$, and also the result
of fitting them to the leading order term in chiral perturbation
theory, $m_{K}m_{\pi}$. The plots show that the data fit the leading
order theory very well.  Of course, here our error bars are larger
than for the $\Delta I = 3/2$ matrix elements and modest chiral
logarithm effects are consistent with the apparent linearity of our
data and our error bars.  It remains an open question as to whether
we can achieve statistical accuracy for the subtracted (8,1)
operators that is the same size as the predicted effects of chiral
logarithms.

Due to the existence of power divergences and the fact that the
residual chiral symmetry breaking effects in divergent amplitudes
are not strictly proportional to $\mres$ , the subtracted matrix
elements do not have to vanish at the chiral limit \cite{Blum:2001xb}.
(Recall that the physical quantity of interest is the slope of the
subtracted matrix elements.)  Rather, at the chiral limit, their
deviation from zero should be an $\mathcal{O} \left(m_{\mathrm{res}}
\right)$ effect. In Figure \ref{fig:Q6 K->pi sub fit}(b), the top line
does not use $m_{\mathrm{res}}$ in the subtraction process
(Eq.\ref{eq:K->pi sub}), while the bottom line does subtraction
with $m_{\mathrm{sea}}$.  As these two lines bracket the zero point
at the chiral limit, the above expectation is verified.

\subsection{PQS vs. PQN}

As we work in the 2+1 flavor partially quenched theory, there are different 
methods to make contractions with the four-quark operators which have a 
singlet piece, {\em i.e.} PQS vs. PQN \cite{Golterman:2006ed,Golterman:2006qv,Laiho:2003uy,Aubin:2006vt}.
The operators $Q_{i}\:\left\{ i=3,4,5,6\right\} $
all have a singlet part in their operator definition,
\begin{equation}
Q_{i}\equiv\left(\bar{s}d\right)_{L}\sum_{q\in\left\{ u,d,s\right\} }\left(\bar{q}q\right)_{L\ \mbox{or}\  R}.\label{eq:PQS op def}
\end{equation}
When evaluating their weak matrix elements on our partially quenched
lattices, we may either keep the singlet
structure of the operator, such that the summation will become a sum
over all valence, sea, and ghost quarks, or we only sum over
the valence quarks as we shall do in full QCD. The former, as we keep 
the singlet property, is
called the PQS (partially quenched singlet) method, while the latter
is called the PQN (partially quenched non-singlet) method. 

\begin{figure}
\begin{centering}
\subfigure[]{\includegraphics[scale=0.7]{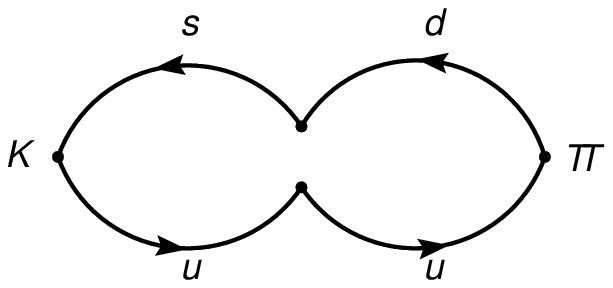}}
\subfigure[]{\includegraphics[scale=0.7]{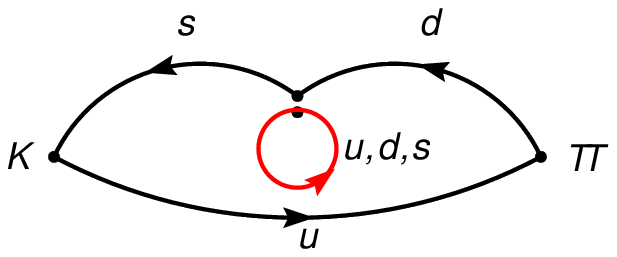}}
\subfigure[]{\includegraphics[scale=0.7]{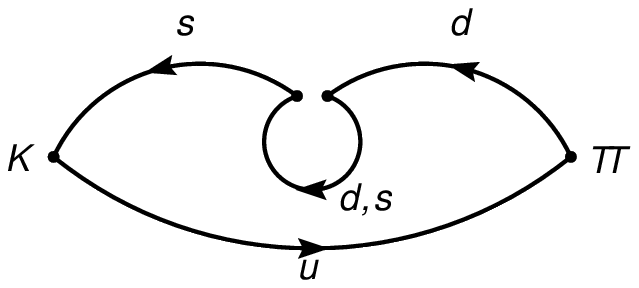}}
\par\end{centering}

\caption{Quark flow diagrams needed to evaluate the weak matrix elements for
$Q_{i}\:\left\{ i=3,4,5,6\right\} $. A connected line represents a trace over 
spin indices. Only diagram (b) has differs
between the PQS and PQN method. With the PQS method, the quark loop (red)
will be constructed from sea quarks, while with the PQN method, it will
be constructed from valence quarks.\label{fig:PQS vs PQN}}
\end{figure}

On the quark level, calculating the weak matrix elements of the
above operators requires the evaluation of the three quark-flow diagrams
shown in Figure \ref{fig:PQS vs PQN}. The only difference between
PQS and PQN is in diagram (b). With the PQS method, the summation
on the quark self-contraction (the red quark loop) will go over all
valence, sea, and ghost quarks (and since valence loops cancel with
ghosts, only the sum over sea remains). With the PQN method,
the sum will go over the valence quarks only.

We have calculated the subtracted matrix elements with both the PQS
and PQN methods (Table \ref{tab:PQS vs PQN}). The comparison of the
slope of the subtracted matrix elements (w.r.t $m_K m_{\pi}$), which
is the leading order coefficient for (8,1) operators, shows small differences
between the two methods. We see that this ambiguity does not affect the chiral fits 
significantly on our 2+1 flavor partially quenched lattices.

\begin{table}
\begin{centering}
\begin{tabular}{c|c|c|c|c}
\hline 
Slope of $\left\langle \pi^{+}\left|Q_{i}\right|K^{+}\right\rangle _{\mathrm{sub}}$&
3&
4&
5&
6\tabularnewline
\hline
PQS&
$0.0009(37)$&
$0.0383(27)$&
$-0.0626(53)$&
$-0.183(13)$\tabularnewline
\hline 
PQN&
$-0.0042(38)$&
$0.0373(34)$&
$-0.0638(55)$&
$-0.165(14)$\tabularnewline
\hline
\end{tabular}
\par\end{centering}

\caption{Comparison between the PQS and PQN methods of the slope 
of the subtracted $K\to\pi$ matrix elements (w.r.t $m_K m_{\pi}$).  \label{tab:PQS vs PQN}}

\end{table}

\section{Non-perturbative Renormalization for $\Delta S=1$ Operators}

\subsection{NPR Procedure}

To convert the bare lattice quantities into the continuum scheme, we have
employed the Rome-Southampton non-perturbative renormalization prescription 
\cite{Martinelli:1994ty}. Following the procedure as described in 
ref. \cite{Blum:2001xb}, we write the NPR formula as
\begin{equation}
O_{i}^{\mathrm{cont,\, ren}}\left(\mu\right)
=
\sum_{j}Z_{ij} \left(\mu\right) 
	\left[ O_{j}^{\mathrm{lat}} + \sum_{k}c_{k}^{j} \left(\mu\right) B_{k}^{\mathrm{lat}} \right]
		 +\mathcal{O}\left(a\right)                            \label{eq:NPR}
\end{equation}
where $O_{i}$ are the dimension 6 operators, and $B_{k}$ are the
lower dimensional operators. The NPR calculation is done on $16^{3}\times32\times16$ lattices \cite{Allton:2007hx} which have been generated
with the same parameters as our $24^3 \times 64 \times 16$ lattices,
except that the lightest dynamical quark mass on the smaller volume is
0.01.

To simplify the computation, we have eliminated $Q_{4}$, $Q_{9}$,
and $Q_{10}$ from the 10-operator basis, since they can be written
as linear combination of the other operators \cite{Blum:2001xb}.
Further, we rotate the remaining 7-operator basis such that each operator
is in a distinct $SU(3)\otimes SU(3)$ representation. The transformation
relation can be found in ref.\cite{Blum:2001xb}. The new set of operators
is denoted as $Q_{i}'$.

\subsection{Resolving the Mixing with Lower Dimensional Operators}

To resolve the mixing between the four quark operator and the lower
dimensional operators, we consider the two most divergent lower dimensional
ones \cite{Blum:2001xb}: 
\begin{align}
B_{1} & \equiv\bar{s}d\nonumber \\
B_{2} & \equiv \bar{s}\left(-\overleftarrow{\slashed{D}}+m_{s}\right)d 
		+ \bar{s}\left(\overrightarrow{\slashed{D}}+m_{d}\right)d.    \label{eq:low dim op}
\end{align}
To compute the corresponding mixing coefficients $c_{1}^{i}$ and $c_{2}^{i}$, we have
used two conditions
\begin{align}
\mathrm{Tr}\left[ \left\langle s \left(p\right) O_{i}^{\mathrm{sub}} \bar{d} \left(p\right) \right\rangle _{\mathrm{amp}} \right] & = 0   \nonumber \\
\mathrm{Tr}\left[ i \slashed{p} \left\langle s \left(p\right) O_{i}^{\mathrm{sub}} \bar{d} \left(p\right) \right\rangle _{\mathrm{amp}} \right] & = 0              \label{eq:low dim eq}
\end{align}
with propagators evaluated at unitary masses ($m_{d}=m_{s}=m_{\mathrm{sea}}$).

\subsection{Resolving the Mixing between Four-Quark Operators}

To determine the mixing coefficient between the 7 operators in the
new basis, we construct a set of external quark combinations $E_{\alpha\beta\gamma\delta}^{j}$
\cite{Blum:2001xb}. Then we impose the renormalization condition that
\begin{equation}
\frac{1}{Z_{q}^{2}} Z^{ki} \Gamma_{\beta\alpha\delta\gamma}^{j} 
	\left\langle O_{i}^{\mathrm{sub}} E_{\alpha\beta\gamma\delta}^{j} \right\rangle _{\mathrm{amp}} 
				= F^{kj}                            \label{eq:NPR cond}
\end{equation}
where $\Gamma_{\beta\alpha\delta\gamma}^{j}$ is the projector corresponding 
to the spin and color structure of the operator $j$,
and there is no sum over $j$ in the above equation. On the r.h.s, $F^{kj}$ is
the free field limit of the matrix 
$\Gamma_{\beta\alpha\delta\gamma}^{j} 
	\left\langle O_{i}^{\mathrm{sub}} E_{\alpha\beta\gamma\delta}^{j} \right\rangle _{\mathrm{amp}}$.
$Z_{q}^{1/2}$ is the quark wavefunction renormalization factor \cite{Blum:2001sr}.

\begin{figure}
\begin{centering}
\includegraphics[bb=60bp 660bp 300bp 720bp,clip,scale=1.5]{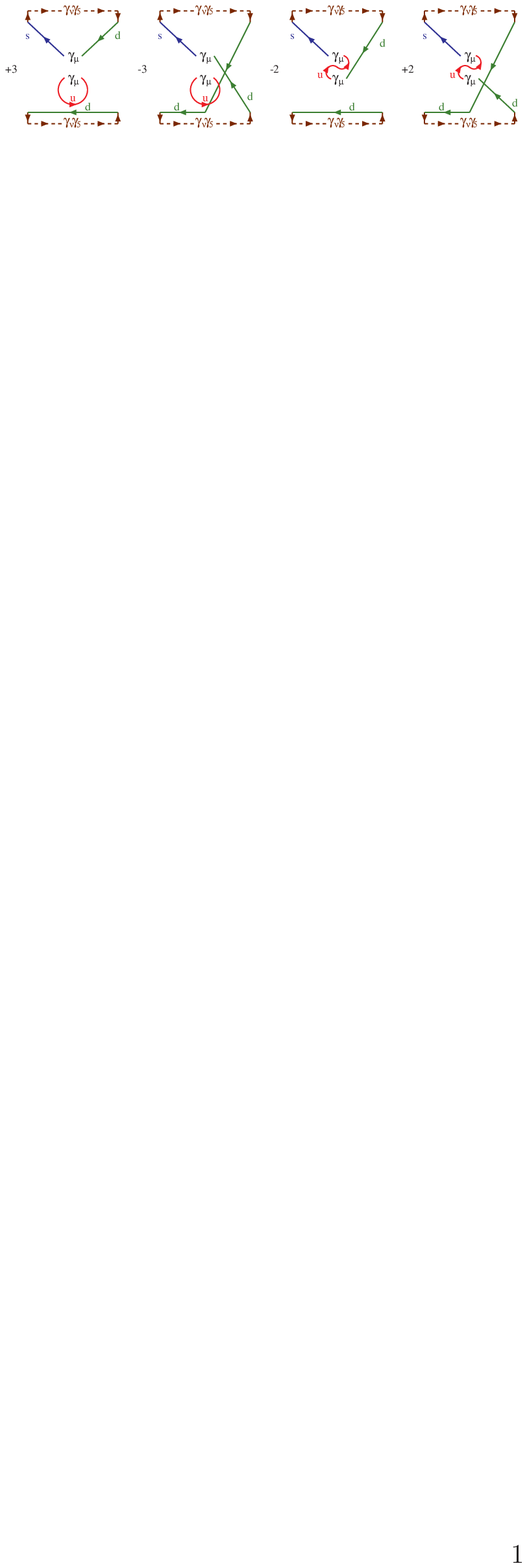}
\par\end{centering}

\caption{4 out of 248 quark flow diagrams with their corresponding factors
involved in the calculation of the mixing coefficients between $Q_{1}'$
and $Q_{1}'$. The solid lines represent quark propagators and each
loop formed by solid and dashed lines represents a trace over spin and 
color indices. \label{fig:npr 4q mix diagram}}
\end{figure}

The calculation of the combination 
$\Gamma_{\beta\alpha\delta\gamma}^{j}
	\left\langle O_{i}^{\mathrm{sub}} E_{\alpha\beta\gamma\delta}^{j} \right\rangle _{\mathrm{amp}}$
requires evaluation of the many quark flow diagrams that can possibly
be
constructed with the relevant operator and the external quark combination.
Figure \ref{fig:npr 4q mix diagram} shows 4 out of 248 quark flow
diagrams involved in the evaluation of renormalization coefficient for $Q_{1}'$.
Each diagram is evaluated at unitary quark mass and non-exceptional external momenta
$p_{1}$ and $p_{2}$ where $p_{1}^{2}=p_{2}^{2}=\left(p_{1}-p_{2}\right)^{2}$
\cite{Blum:2001xb}.

\begin{figure}
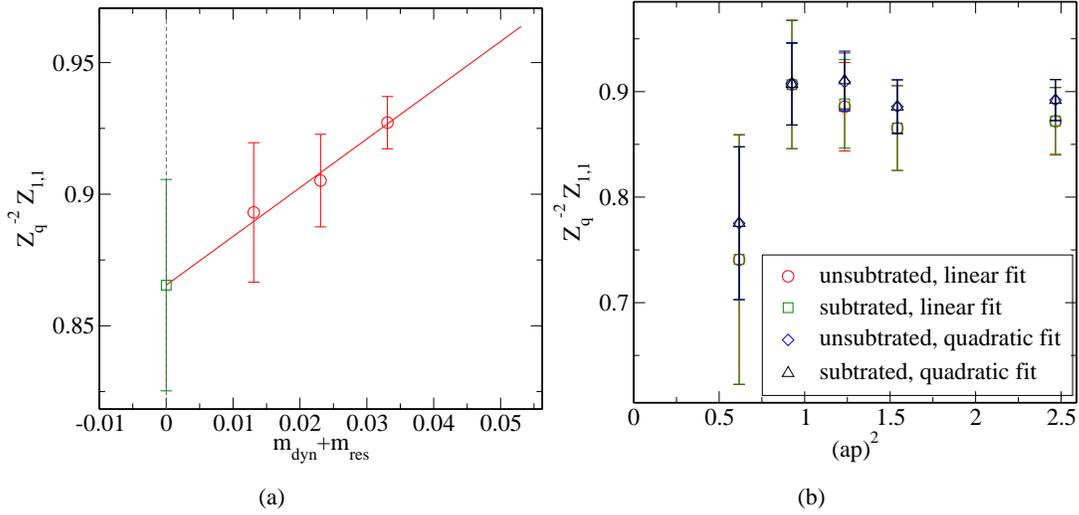

\begin{centering}
\subfigure[]{\includegraphics[clip,scale=0.6]{raw/FMinv_subtd_linfit_Op_O1_Ext_O1_ap2_160}}
\subfigure[]{\includegraphics[clip,scale=0.6]{raw/FMinv_scale_Op_O1_Ext_O1}}
\par\end{centering}

\caption{Evaluation of the mixing coefficients between four quark operators.
Figure (a) shows the linear mass fit of the mixing coefficient $Z_{q}^{-2}Z_{1,1}$.
Figure (b) shows the final value of $Z_{q}^{-2}Z_{1,1}$ at 5 different
momentum scales where we have sufficient pairs of non-exceptional
momenta $\left(p_{1},\, p_{2}\right)$ to evaluate the quark flow
diagrams. Also in Figure (b), we show a comparison between linear and quadratic mass fits,
as well as a comparison between the values before and after the subtraction of
$c_{2}^{i}$. It is clear that the differences are not significant.
\label{fig:npr result}}
\end{figure}

After the evaluation of the entire matrix of mixing coefficients, we remove the
coefficients that are both theoretically suppressed and statistically
zero, and then fit the remaining matrix to a unitary mass dependence
(Figure \ref{fig:npr result}(a)). Since one of the lower
dimensional mixing coefficients, $c_{1}^{i}$, is expected to have the
mass dependence
$m/a^{2}$ at leading order, this step will automatically remove
$c_{1}^{i}$. Then the subtraction of the other, $c_{2}^{i}$, is performed.
Due to its small magnitude, this subtraction does not make big differences in the final
result (Figure \ref{fig:npr result}(b)).

\section{Conclusions}
We have discussed the status of measurements of $K \to \pi$ and $K
\to 0$ weak matrix elements on $24^3 \times 64$, 2+1 flavor dynamical
domain wall lattices by the RBC and UKQCD collaborations.  We are
working at much smaller masses and larger volumes than earlier
quenched calculations and, with the 2+1 flavor ensembles, we are
free of quenching errors.  We find good statistical errors and long
plateaus for our measured matrix elements and the subtraction of
power divergent operator mixing contributions to (8,1) operators
works as expected.  We find that the ambiguity between the PQS and
PQN methods is removed.  The non-perturbative renormalization of our
operators, following the Rome-Southampton prescription, is essentially
complete.  We await the completion of the NLO $PQ\chi PT$ formulae
for these matrix elements, and preliminary fits are underway.  The
accuracy and stability of these fits will play a large role in
determining our final errors.

\section*{Acknowledgment}

This project is being carried out in collaboration with members of
the RBC and UKQCD collaborations.  We thank our colleagues in the
RBC and UKQCD collaborations for the development and support of the
hardware and software infrastructure which was essential to this
work.  In addition we acknowledge the University of Edinburgh,
PPARC, RIKEN, BNL and the U.S.\ DOE for providing the facilities
on which this work was performed.

\bibliographystyle{apsrev}
\bibliography{samref}

\begin{thebibliography}{10}
\expandafter\ifx\csname bibnamefont\endcsname\relax
  \def\bibnamefont#1{#1}\fi
\expandafter\ifx\csname bibfnamefont\endcsname\relax
  \def\bibfnamefont#1{#1}\fi
\expandafter\ifx\csname url\endcsname\relax
  \def\url#1{\texttt{#1}}\fi
\expandafter\ifx\csname urlprefix\endcsname\relax\def\urlprefix{URL }\fi
\expandafter\ifx\csname bibinfo\endcsname\relax \def\bibinfo#1#2{#2}\fi
\expandafter\ifx\csname eprint\endcsname\relax \def\eprint#1{#1}\fi

\bibitem{Lellouch:2000pv}
\bibinfo{author}{\bibfnamefont{L.}~\bibnamefont{Lellouch}} \bibnamefont{and}
  \bibinfo{author}{\bibfnamefont{M.}~\bibnamefont{Luscher}},
  \bibinfo{journal}{Commun. Math. Phys.} \textbf{\bibinfo{volume}{219}},
  \bibinfo{pages}{31} (\bibinfo{year}{2001}), \eprint{hep-lat/0003023}.

\bibitem{Kim:2004sk}
\bibinfo{author}{\bibfnamefont{C.}~\bibnamefont{Kim}}
  \bibinfo{note}{UMI-31-47246}.

\bibitem{Bernard:1985wf}
\bibinfo{author}{\bibfnamefont{C.~W.} \bibnamefont{Bernard}},
  \bibinfo{author}{\bibfnamefont{T.}~\bibnamefont{Draper}},
  \bibinfo{author}{\bibfnamefont{A.}~\bibnamefont{Soni}},
  \bibinfo{author}{\bibfnamefont{H.~D.} \bibnamefont{Politzer}},
  \bibnamefont{and} \bibinfo{author}{\bibfnamefont{M.~B.} \bibnamefont{Wise}},
  \bibinfo{journal}{Phys. Rev.} \textbf{\bibinfo{volume}{D32}},
  \bibinfo{pages}{2343} (\bibinfo{year}{1985}).

\bibitem{Blum:2001xb}
\bibinfo{author}{\bibfnamefont{T.}~\bibnamefont{Blum}} \emph{et~al.}
  (\bibinfo{collaboration}{RBC}), \bibinfo{journal}{Phys. Rev.}
  \textbf{\bibinfo{volume}{D68}}, \bibinfo{pages}{114506}
  (\bibinfo{year}{2003}), \eprint{hep-lat/0110075}.

\bibitem{Noaki:2001un}
\bibinfo{author}{\bibfnamefont{J.~I.} \bibnamefont{Noaki}} \emph{et~al.}
  (\bibinfo{collaboration}{CP-PACS}), \bibinfo{journal}{Phys. Rev.}
  \textbf{\bibinfo{volume}{D68}}, \bibinfo{pages}{014501}
  (\bibinfo{year}{2003}), \eprint{hep-lat/0108013}.

\bibitem{Golterman:2001qj}
\bibinfo{author}{\bibfnamefont{M.}~\bibnamefont{Golterman}} \bibnamefont{and}
  \bibinfo{author}{\bibfnamefont{E.}~\bibnamefont{Pallante}},
  \bibinfo{journal}{JHEP} \textbf{\bibinfo{volume}{10}}, \bibinfo{pages}{037}
  (\bibinfo{year}{2001}), \eprint{hep-lat/0108010}.

\bibitem{Golterman:2002us}
\bibinfo{author}{\bibfnamefont{M.}~\bibnamefont{Golterman}} \bibnamefont{and}
  \bibinfo{author}{\bibfnamefont{E.}~\bibnamefont{Pallante}},
  \bibinfo{journal}{Phys. Rev.} \textbf{\bibinfo{volume}{D69}},
  \bibinfo{pages}{074503} (\bibinfo{year}{2004}), \eprint{hep-lat/0212008}.

\bibitem{Golterman:2006ed}
\bibinfo{author}{\bibfnamefont{M.}~\bibnamefont{Golterman}} \bibnamefont{and}
  \bibinfo{author}{\bibfnamefont{E.}~\bibnamefont{Pallante}},
  \bibinfo{journal}{Phys. Rev.} \textbf{\bibinfo{volume}{D74}},
  \bibinfo{pages}{014509} (\bibinfo{year}{2006}), \eprint{hep-lat/0602025}.

\bibitem{Antonio:2006zza}
\bibinfo{author}{\bibfnamefont{D.~J.} \bibnamefont{Antonio}} \emph{et~al.}
  (\bibinfo{collaboration}{RBC and UKQCD}), \bibinfo{journal}{PoS}
  \textbf{\bibinfo{volume}{LAT2006}}, \bibinfo{pages}{188}
  (\bibinfo{year}{2006}).

\bibitem{Antonio:2007pb}
\bibinfo{author}{\bibfnamefont{D.~J.} \bibnamefont{Antonio}} \emph{et~al.}
  (\bibinfo{collaboration}{RBC and UKQCD})  (\bibinfo{year}{2007}),
  \eprint{hep-ph/0702042}.

\bibitem{Lin:2007lat}
\bibinfo{author}{\bibfnamefont{M.}~\bibnamefont{Lin}} \bibnamefont{and}
  \bibinfo{author}{\bibfnamefont{E.~E.} \bibnamefont{Scholz}}
  (\bibinfo{collaboration}{RBC and UKQCD}), \bibinfo{journal}{these
  proceedings} .

\bibitem{Antonio:2007lat}
\bibinfo{author}{\bibfnamefont{D.~J.} \bibnamefont{Antonio}} \bibnamefont{and}
  \bibinfo{author}{\bibfnamefont{S.~D.} \bibnamefont{Cohen}}
  (\bibinfo{collaboration}{RBC and UKQCD}), \bibinfo{journal}{these
  proceedings} .

\bibitem{Aubin:2007xpt}
\bibinfo{author}{\bibfnamefont{C.}~\bibnamefont{Aubin}},
  \bibinfo{author}{\bibfnamefont{J.}~\bibnamefont{Laiho}}, \bibnamefont{and}
  \bibinfo{author}{\bibfnamefont{S.}~\bibnamefont{Li}}, \bibinfo{journal}{in
  preparation} .

\bibitem{Golterman:2006qv}
\bibinfo{author}{\bibfnamefont{M.}~\bibnamefont{Golterman}} \bibnamefont{and}
  \bibinfo{author}{\bibfnamefont{E.}~\bibnamefont{Pallante}},
  \bibinfo{journal}{PoS} \textbf{\bibinfo{volume}{LAT2006}},
  \bibinfo{pages}{090} (\bibinfo{year}{2006}), \eprint{hep-lat/0610005}.

\bibitem{Laiho:2003uy}
\bibinfo{author}{\bibfnamefont{J.}~\bibnamefont{Laiho}} \bibnamefont{and}
  \bibinfo{author}{\bibfnamefont{A.}~\bibnamefont{Soni}},
  \bibinfo{journal}{Phys. Rev.} \textbf{\bibinfo{volume}{D71}},
  \bibinfo{pages}{014021} (\bibinfo{year}{2005}), \eprint{hep-lat/0306035}.

\bibitem{Aubin:2006vt}
\bibinfo{author}{\bibfnamefont{C.}~\bibnamefont{Aubin}} \emph{et~al.},
  \bibinfo{journal}{Phys. Rev.} \textbf{\bibinfo{volume}{D74}},
  \bibinfo{pages}{034510} (\bibinfo{year}{2006}), \eprint{hep-lat/0603025}.

\bibitem{Martinelli:1994ty}
\bibinfo{author}{\bibfnamefont{G.}~\bibnamefont{Martinelli}},
  \bibinfo{author}{\bibfnamefont{C.}~\bibnamefont{Pittori}},
  \bibinfo{author}{\bibfnamefont{C.~T.} \bibnamefont{Sachrajda}},
  \bibinfo{author}{\bibfnamefont{M.}~\bibnamefont{Testa}}, \bibnamefont{and}
  \bibinfo{author}{\bibfnamefont{A.}~\bibnamefont{Vladikas}},
  \bibinfo{journal}{Nucl. Phys.} \textbf{\bibinfo{volume}{B445}},
  \bibinfo{pages}{81} (\bibinfo{year}{1995}), \eprint{hep-lat/9411010}.

\bibitem{Allton:2007hx}
\bibinfo{author}{\bibfnamefont{C.}~\bibnamefont{Allton}} \emph{et~al.}
  (\bibinfo{collaboration}{RBC and UKQCD}), \bibinfo{journal}{Phys. Rev.}
  \textbf{\bibinfo{volume}{D76}}, \bibinfo{pages}{014504}
  (\bibinfo{year}{2007}), \eprint{hep-lat/0701013}.

\bibitem{Blum:2001sr}
\bibinfo{author}{\bibfnamefont{T.}~\bibnamefont{Blum}} \emph{et~al.},
  \bibinfo{journal}{Phys. Rev.} \textbf{\bibinfo{volume}{D66}},
  \bibinfo{pages}{014504} (\bibinfo{year}{2002}), \eprint{hep-lat/0102005}.

\end{thebibliography}

\end{document}